\documentstyle[12pt]{article}
\input amssym.def
\input amssym.tex

\newcommand{\wiq}{\widetilde{q}}
\newcommand{\wir}{\widetilde{r}}
\newcommand{\whq}{\widehat{q}}
\newcommand{\whr}{\widehat{r}}
\newcommand{\bq}{{\bf q}}
\newcommand{\br}{{\bf r}}
\newcommand{\wbq}{\widetilde{\bf q}}
\newcommand{\wbr}{\widetilde{\bf r}}

\begin{document}

\renewcommand{\theequation}{\thesection.\arabic{equation}}

%
%

\begin{titlepage}
 {\LARGE
  \begin{center} 
A note on the integrable discretization of the nonlinear
Schr\"odinger equation 
  \end{center}
 }

\vspace{1.5cm}

\begin{flushleft}{\large Yuri B. SURIS}\end{flushleft} \vspace{1.0cm}
Centre for Complex Systems and Visualization, University of Bremen,\\
Universit\"atsallee 29, 28359 Bremen, Germany\\
e-mail: suris @ cevis.uni-bremen.de 

\vspace{2.0cm}

{\small {\bf Abstract.} We revisit integrable discretizations 
for the nonlinear Schr\"odinger equation due to Ablowitz and Ladik. 
We demonstrate how their main drawback, the non-locality, can be 
overcome. Namely, we factorize the non-local difference scheme into the 
product of local ones. This must improve the performance of the scheme
in the numerical computations dramatically. Using the equivalence
of the Ablowitz--Ladik and the relativistic Toda hierarchies, we find
the interpolating Hamiltonians for the local schemes and show how to
solve them in terms of matrix factorizations.}
\end{titlepage}

\setcounter{equation}{0}
\section{Introduction}
In \cite{S1}--\cite{S4} the author pushed forward a new method of 
finding integrable discretizations for integrable differential equations,
based on the notion of $r$--matrix hierarchies and the related mathematical
apparatus. The main idea of this approach is to seak for integrable 
discretizations in the same hierarchies where their continuous counterparts
live.

In fact, this method is not quite new. The first integrable discretizations
which can be treated as an application of this method go back as far as to
the year 1976, to the work of Ablowitz and Ladik \cite{AL2}. In the present 
note we revisit the Ablowitz--Ladik discretizations, improving them both
from the aesthetical (theoretical) and the practical (computational) point of
view.

In \cite{AL1} Ablowitz and Ladik proposed the following remarkable system 
of ordinary differential equations:
\begin{eqnarray}\label{AL}
\dot{q}_k & = & q_{k+1}-2q_k+q_{k-1}-q_kr_k(q_{k+1}+q_{k-1})\nonumber\\ \\
\dot{r}_k & = & -r_{k+1}+2r_k-r_{k-1}+q_kr_k(r_{k+1}+r_{k-1}).\nonumber
\end{eqnarray}
It may be considered either on an infinite lattice ($k\in {\Bbb Z}$) under
the boundary conditions of a rapid decay ($|q_k|, |r_k|\to 0$ as 
$k\to\pm\infty$),
or on a finite lattice ($1\le k\le N$) under the periodic boundary conditions 
($q_0\equiv q_N$, $r_0\equiv r_N$, $q_{N+1}\equiv q_1$, $r_{N+1}\equiv r_1$).
In any case we shall denote by $\bq$ ($\br$) the (infinite- or 
finite-dimensional) vector with the components $q_k$ (resp. $r_k$).
 
In \cite{AL1} the system (\ref{AL}) appeared as a space discretization of the 
following system of partial differential equations:
\begin{equation}\label{cNLS}
q_t=q_{xx}-2q^2r,\qquad r_t=-r_{xx}+2qr^2
\end{equation}
(to perform the corresponding continuous limit, one has first to rescale
in (\ref{AL}) $t\mapsto \epsilon^{-2}t$, $q_k\mapsto \epsilon q_k$, 
$r_k\mapsto\epsilon r_k$, and then to send $\epsilon\to 0$).

It is important to notice that upon the change of the independent variable
$t\mapsto it$, $i=\sqrt{-1}$, the system (\ref{cNLS}) allows a reduction
\begin{equation}\label{red}
r=\pm q^*,
\end{equation}
leading to the nonlinear Schr\"odinger equation 
\begin{equation}\label{NLS}
-iq_t=q_{xx}\mp 2|q|^2q.
\end{equation}
(In (\ref{red}) and below we use the asterisque $^*$ to denote the complex 
conjugation). The same reduction is admissible also by the Ablowitz--Ladik
system (\ref{AL}), leading to
\begin{equation}\label{AL real}
-i\dot{q}_k=q_{k+1}-2q_k+q_{k-1}\mp|q_k|^2(q_{k+1}+q_{k-1}).
\end{equation}

Ablowitz and Ladik found also a commutation representation for the system
(\ref{AL}) -- a semi-discrete version of the zero--curvature representation:
\begin{equation}\label{AL Lax}
\dot{L}_k=M_{k+1}L_k-L_kM_k
\end{equation}
with $2\times 2$ matrices $L_k$, $M_k$ depending on the variables $\bq, \br$ 
and on the additional (spectral) parameter $\lambda$:
\begin{equation}\label{AL Lk}
L_k=L_k(\bq,\br)=\left(
  \begin{array}{cc} \lambda & q_k \\ \\ r_k & \lambda^{-1}\end{array}
     \right),
\end{equation}
\begin{equation}\label{AL Mk}
M_k=M_k(\bq,\br)=\left(
   \begin{array}{cc} \lambda^2-1-q_kr_{k-1} & \lambda q_k-\lambda^{-1}q_{k-1}\\
    \\ \lambda r_{k-1}- \lambda^{-1} r_k & -\lambda^{-2}+1+ q_{k-1}r_k 
   \end{array}
     \right).
\end{equation}
Note that the linear problem associated with the matrix $L_k$,
\begin{equation}\label{d lin pr}
\Psi_{k+1}=L_k\Psi_k,
\end{equation}
is a discretization of the linear Zakharov--Shabat problem, associated with
the system (\ref{cNLS}),
\begin{equation}\label{lin pr}
\Psi_x=\left(\begin{array}{cc}i\zeta & q \\ \\ r & -i\zeta\end{array}\right)\Psi.
\end{equation}

In \cite{AL2} Ablowitz and Ladik made also the next step in discretizing the
system (\ref{cNLS}): they constructed  a family of time discretizations of the 
system (\ref{AL}). Although it was not stressed very explicitly, this time
their approach to discretization was fundamentally different: they did not
modify the linear problem (\ref{d lin pr}) any more, restricting themselves
with a choice of a suitable (discrete)--time evolution of the wave function
$\Psi_k$. Hence the basic feature of the time--discretizations in \cite{AL2} 
is following: they admit a discrete analog of the zero--curvature representation,
\begin{equation}\label{d zero curv}
\widetilde{L}_kV_k=V_{k+1}L_k
\end{equation}
with the same matrix $L_k$ as the underlying continuous time system. (In 
(\ref{d zero curv}) and below we use the tilde to denote the $h$--shift
in the discrete time $h{\Bbb Z}$). In a more modern language, the maps
generated by the discretizations in \cite{AL2} belong to the same integrable 
hierarchy as the continuous time system (\ref{AL}).

The results of \cite{AL2} may be formulated as follows. 

{\bf Proposition 0.} {\it Let the matrix $L_k$ be given by} (\ref{AL Lk}),
{\it and let the entries of the matrix 
\begin{equation}\label{dAL Vk}
V_k=\left(\begin{array}{cc}{\cal A}_k & {\cal B}_k\\ \\
{\cal C}_k & {\cal D}_k\end{array}\right)
\end{equation}
have the following $\lambda$--dependence:
\[
{\cal A}_k=1+h\lambda^2{\cal A}_k^{(2)}+h{\cal A}_k^{(0)}+
h\lambda^{-2}{\cal A}_k^{(-2)},
\]
\[
{\cal D}_k=1+h\lambda^2{\cal D}_k^{(2)}+h{\cal D}_k^{(0)}+
h\lambda^{-2}{\cal D}_k^{(-2)},
\]
\[
{\cal B}_k=h\lambda{\cal B}_k^{(1)}+h\lambda^{-1}{\cal B}_k^{(-1)},
\]
\[
{\cal C}_k=h\lambda{\cal C}_k^{(1)}+h\lambda^{-1}{\cal C}_k^{(-1)}.
\]
Then the discrete zero--curvature equation} (\ref{d zero curv}) {\it implies
the following expressions:
\begin{eqnarray}
{\cal A}_k&=&1-h\alpha_0+h\alpha_+(\lambda^2-A_k)+h(\alpha_-\lambda^{-2}-
\delta_-\wiq_{k}r_{k-1})\Lambda_k,\label{dAL cAk}\\
{\cal D}_k&=&1+h\delta_0-h\delta_+(\lambda^{-2}-D_k)-h(\delta_-\lambda^2-
\alpha_-q_{k-1}\wir_k)\Lambda_k,\label{dAL cDk}\\
{\cal B}_k&=&h(\alpha_+\lambda q_k-\delta_+\lambda^{-1}\wiq_{k-1})+
h(\delta_-\lambda\wiq_k-\alpha_-\lambda^{-1}q_{k-1})\Lambda_k,\label{dAL cBk}\\
{\cal C}_k&=&h(\alpha_+\lambda\wir_{k-1}-\delta_+\lambda^{-1}r_k)+
h(\delta_-\lambda r_{k-1}-\alpha_-\lambda^{-1}\wir_k)\Lambda_k,\label{dAL cCk}
\end{eqnarray}
together with the equations of motion:
\begin{eqnarray}\label{dAL}
(\wiq_k-q_k)/h & = & 
\alpha_+q_{k+1}-\alpha_0q_k-\delta_0\wiq_k+\delta_+\wiq_{k-1}-
(\alpha_+q_kA_{k+1}+\delta_+\wiq_kD_k)\nonumber\\
 & & +(\delta_-\wiq_{k+1}+\alpha_-q_{k-1})(1-\wiq_k\wir_k)\Lambda_k\nonumber\\ 
\\
(\wir_k-r_k)/h & = & 
-\delta_+r_{k+1}+\delta_0r_k+\alpha_0\wir_k-\alpha_+\wir_{k-1}+
(\delta_+r_kD_{k+1}+\alpha_+\wir_kA_k)\nonumber\\
 & & -(\alpha_-\wir_{k+1}+\delta_-r_{k-1})(1-\wiq_k\wir_k)\Lambda_k\nonumber
\end{eqnarray}
Here $\alpha_0$, $\alpha_+$, $\alpha_-$, $\delta_0$, $\delta_+$, $\delta_-$
are constants, and the functions $A_k$, $D_k$, $\Lambda_k$ satisfy the 
following difference relations}
\begin{equation}\label{dif rel Ak}
A_{k+1}-A_k = q_{k+1}r_k-\wiq_k\wir_{k-1},
\end{equation}
\begin{equation}\label{dif rel Dk}
D_{k+1}-D_k = q_kr_{k+1}-\wiq_{k-1}\wir_k,
\end{equation}
\begin{equation}\label{dif rel Lamk}
\Lambda_{k+1}(1-q_kr_k)=\Lambda_k(1-\wiq_k\wir_k).
\end{equation}

{\bf Remark.} In the case of the rapidly decreasing boundary conditions
there exists a canonical way to single out certain solutions of the
difference equations (\ref{dif rel Ak})--(\ref{dif rel Lamk}) above,
namely, by the conditions
\[
A_k,\;D_k\to 0,\quad \Lambda_k\to 1\quad {\rm as} \quad k\to\pm\infty,
\]
which results in
\begin{equation}\label{dAL Ak}
A_k = q_kr_{k-1}+\sum_{j=-\infty}^{k-1} (q_jr_{j-1}-\wiq_j\wir_{j-1}),
\end{equation}
\begin{equation}\label{dAL Dk}
D_k = q_{k-1}r_k+\sum_{j=-\infty}^{k-1} (q_{j-1}r_j-\wiq_{j-1}\wir_j),
\end{equation}
\begin{equation}\label{dAL Lamk}
\Lambda_k=\prod_{j=-\infty}^{k-1} \frac{1-\wiq_j\wir_j}{1-q_jr_j}.
\end{equation}
In the case of the periodic boundary conditions to choose a particular
solution one can use the same formulas with the sums and product starting 
from $j=0$ instead of $j=-\infty$.

The numbers $\alpha_0$, $\delta_0$, $\alpha_-$, $\delta_-$, playing the role
of constants of summation, are defined as soon as certain solutions $A_k$, 
$D_k$, $\Lambda_k$ have been fixed.

The formulas (\ref{dAL}), (\ref{dAL Ak})--(\ref{dAL Lamk}) define the map 
which we shall denote
\[
{\cal T}_{\rm AL}(h;\alpha_0,\alpha_+,\alpha_-;
\delta_0,\delta_+,\delta_-):\quad(\bq,\br)\mapsto(\wbq,\wbr)
\]
It is important to notice that for the
pure imaginary values of $h$ this map obviously allows the reduction
\[
\br=\pm\bq^*,
\]
provided $A_k$, $D_k$, $\Lambda_k$ are chosen as in 
(\ref{dAL Ak})--(\ref{dAL Lamk}), and
\begin{equation}\label{del = al conj}
\delta_0=\alpha_0^*,\quad\delta_+=\alpha_+^*,\quad\delta_-=\alpha_-^*,
\end{equation}
so that that there remains a three--parameter family of difference schemes
satisfying this condition.

The expressions (\ref{dAL Ak})--(\ref{dAL Lamk})  serve as 
a source of a non-locality of the difference scheme, which is its major
drawback. This feature makes any numerical realization of the numerical
scheme extremely time--consuming.

The numerical experiments reported in \cite{AT3} showed that even despite 
this drawback the Ablowitz--Ladik difference schemes are the best among
the class of finite difference methods, being surpassed only by
certain spectral numerical methods.

In the present note we shall demonstrate how to factorize the non-local scheme
(\ref{dAL}) into the product of very simple (in particular, local) schemes, 
which surely can speed up the performance of this scheme considerably.

\setcounter{equation}{0}
\section{Ablowitz--Ladik hierarchy \newline and its simplest flows}

From the modern point of view, the Ablowitz--Ladik system (\ref{AL}) is
a representative of a whole hierarchy of commuting Hamiltonian flows. 
Considering, for notational simplicity, the finite dimensional case, we define 
the Poisson bracket on the space ${\Bbb R}^{2N}(\bq,\br)$ by the formula
\begin{equation}\label{AL PB}
\{q_k,r_j\}=(1-q_kr_k)\delta_{jk},\quad \{q_k,q_j\}=\{r_k,r_j\}=0.
\end{equation}
The Hamiltonians of the commuting flows are the coefficients in the Laurent
expansion of the trace ${\rm tr}\; T_N(\bq,\br,\lambda)$ where $T_N$ is the 
monodromy matrix
\begin{equation}\label{AL monodromy}
T_N=L_N\cdot L_{N-1}\cdot\ldots\cdot L_2\cdot L_1,
\end{equation}
supplied by the function
\begin{equation}\label{AL H0}
H_0(\bq,\br)=\log\det T_N=\sum_{k=1}^N\log(1-q_kr_k).
\end{equation}
The involutivity of all integrals of motion follows from the fundamental
$r$--matrix relation:
\begin{equation}\label{r mat PB}
\{L(\lambda)\stackrel{\bigotimes}{,}L(\mu)\}=
[L(\lambda)\otimes L(\mu),\rho(\lambda,\mu)],
\end{equation}
where
\begin{equation}\label{AL r mat}
\rho(\lambda,\mu)=\left(\begin{array}{cccc}
\frac{1}{2}\frac{\lambda^2+\mu^2}{\lambda^2-\mu^2} & 0 & 0 & 0 \\
0 & \frac{1}{2} & \frac{\lambda\mu}{\lambda^2-\mu^2} & 0 \\
0 & \frac{\lambda\mu}{\lambda^2-\mu^2} & -\frac{1}{2} & 0 \\
0 & 0 & 0 & \frac{1}{2}\frac{\lambda^2+\mu^2}{\lambda^2-\mu^2}
\end{array}\right).
\end{equation}

It is easy to see that the following two functions belong to the involutive
family generated by ${\rm tr}\; T_N$:
\begin{equation}\label{AL H pm}
H_+(\bq,\br)=\sum_{k=1}^N q_{k+1}r_k, \quad 
H_-(\bq,\br)=\sum_{k=1}^N q_kr_{k+1}.
\end{equation}

The corresponding Hamiltonian flows are described by the differential equations
\begin{equation}\label{AL flow+}
{\cal F}_+:\quad \dot{q}_k=q_{k+1}(1-q_kr_k),\quad \dot{r}_k=-r_{k-1}(1-q_kr_k),
\end{equation}
\begin{equation}\label{AL flow-}
{\cal F}_-:\quad \dot{q}_k=q_{k-1}(1-q_kr_k),\quad \dot{r}_k=-r_{k+1}(1-q_kr_k),
\end{equation}
The flow generated by the Hamiltonian function (\ref{AL H0}) is described, up
to the factor 2, by the differential equations
\begin{equation}\label{AL flow0}
{\cal F}_0:\quad \dot{q}_k=-2q_k,\quad \dot{r}_k=2r_k.
\end{equation}

The Ablowitz--Ladik flow proper is an obvious linear combination of these
more fundamental and simple flows. According to the general theory \cite{FT}, 
each of the flows ${\cal F}_{\pm}$, ${\cal F}_0$ is described by the 
zero--curvature representation (\ref{d zero curv}) with the same matrix $L_k$, 
but with different matrices $M_k$. The corresponding matrices $M_k$ are given 
by:
\begin{equation}\label{AL M+}
{\cal F}_+:\qquad M^{(+)}_k=\left(\begin{array}{cc}
\lambda^2-q_kr_{k-1} & \lambda q_k\\ \\ \lambda r_{k-1} & 0 \end{array}\right).
\end{equation}
\begin{equation}\label{AL M-}
{\cal F}_-:\qquad M^{(-)}_k=\left(\begin{array}{cc}
0 & -\lambda^{-1} q_{k-1}\\ \\ -\lambda^{-1} r_k & -\lambda^{-2}+q_{k-1}r_k 
\end{array}\right).
\end{equation}
\begin{equation}\label{AL M0}
{\cal F}_0:\qquad M^{(0)}_k=\left(\begin{array}{cc}-1 & 0\\ \\ 
                             0 & 1 \end{array}\right).
\end{equation}

\setcounter{equation}{0}
\section{Local discretizations for ${\cal F}_{\pm}$}

We demonstrate now an unexpected fact. Namely, let only one of the four 
parameters $\alpha_+$, $\delta_+$, $\alpha_-$, $\delta_-$ of the 
Ablowitz--Ladik scheme not vanish (so that the resulting scheme 
approximates one of the flows ${\cal F}_{\pm}$ rather than the original system 
(\ref{AL})). Then it is possible to render the scheme {\it local}. 
This results in four integrable maps $(\bq,\br)\mapsto (\wbq,\wbr)$ which
are described by local equations of motion and belong to the Ablowitz--Ladik 
hierachy, i.e. admit commutation representations with the matrix $L_k$ from
(\ref{AL Lk}). The commutation representations for the maps given in the
following four Propositions could be proved by an easy and
direct check, but we prefere to trace back the relations between our maps
and the original formulation by Ablowitz and Ladik.

{\bf Proposition 1.} {\it The Ablowitz--Ladik scheme {\rm (\ref{dAL})} with 
the parameters
\[
\alpha_0=\delta_0=\alpha_-=\delta_-=\delta_+=0, \quad \alpha_+=1
\]
is equivalent to the following map:
\begin{equation}\label{dAL flow+}
{\cal T}_+(h):\left\{\begin{array}{l}
(\wiq_k-q_k)/h=q_{k+1}(1-q_k\wir_k),\\ \\
(\wir_k-r_k)/h=-\wir_{k-1}(1-q_k\wir_k)\end{array}\right.
\end{equation}
approximating the flow ${\cal F}_+$. This map has the commutation 
representation
\[
{\cal T}_+(h):\quad \widetilde{L}_kV_k^{(+)}=V_{k+1}^{(+)}L_k
\]
with the matrix}
\begin{equation}\label{dAL Vk+}
V_k^{(+)}=V_k^{(+)}(\bq,\wbr,h)=\left(
\begin{array}{cc}1+h\lambda^2-hq_k\wir_{k-1} & h\lambda q_k\\ \\
h\lambda \wir_{k-1} & 1\end{array}\right).
\end{equation}

{\bf Proof.} According to the Proposition 0, the matrix $V_k^{(+)}$ for the
scheme ${\cal T}_{\rm AL}(h;0,1,0;0,0,0)$ has the form
\[
V_k^{(+)}=\left(
\begin{array}{cc}1+h\lambda^2-hA_k & h\lambda q_k\\ \\
h\lambda \wir_{k-1} & 1\end{array}\right),
\]
while the equations of motion read:
\begin{equation}\label{proof1 aux1}
(\wiq_k-q_k)/h=q_{k+1}-q_kA_{k+1},\quad (\wir_k-r_k)/h=-\wir_{k-1}+\wir_kA_k.
\end{equation}
Here $A_k$ is the solution of the difference relation
\begin{equation}\label{proof1 aux2}
A_{k+1}-A_k=q_{k+1}r_k-\wiq_k\wir_{k-1}
\end{equation}
tending to $0$ as $k\to\pm\infty$ in the case of the rapidly decaying boundary
conditions. The Proposition will be demonstrated if we prove the following
formula for $A_k$:
\begin{equation}\label{proof1 aux3}
A_k=q_k\wir_{k-1}
\end{equation}
which comes on the place of the non-local expression (\ref{dAL Ak}). To
this end multiply the first equation in (\ref{proof1 aux1}) by $\wir_{k-1}$,
the second one by $q_{k+1}$, and add the two resulting equations:
\[
\wiq_k\wir_{k-1}-q_{k+1}r_k+q_{k+1}\wir_k-q_k\wir_{k-1}=
hq_{k+1}\wir_kA_k-hq_k\wir_{k-1}A_{k+1}.
\]
Using (\ref{proof1 aux2}) we obtain:
\[
-A_{k+1}+A_k+q_{k+1}\wir_k-q_k\wir_{k-1}=
hq_{k+1}\wir_kA_k-hq_k\wir_{k-1}A_{k+1},
\]
which is equivalent to
\[
\frac{1-hA_k}{1-hq_k\wir_{k-1}}=\frac{1-hA_{k+1}}{1-q_{k+1}\wir_k}={\rm const}.
\]
As the both quantities $A_k$, $q_k\wir_{k-1}$ tend to $0$ by $k\to\pm\infty$,
this constant has to be equal to 1, which ends the proof.

{\bf Proposition 2.} {\it The Ablowitz--Ladik scheme {\rm (\ref{dAL})} with 
the parameters
\[
\alpha_0=\delta_0=\alpha_-=\delta_-=\alpha_+=0, \quad \delta_+=1
\]
is equivalent to the following map:
\begin{equation}\label{dAL flow-}
{\cal T}_-(h):\left\{\begin{array}{l}
(\wiq_k-q_k)/h=\wiq_{k-1}(1-\wiq_kr_k),\\ \\
(\wir_k-r_k)/h=-r_{k+1}(1-\wiq_kr_k)\end{array}\right.
\end{equation}
approximating the flow ${\cal F}_-$. This map has the commutation 
representation
\[
{\cal T}_-(h):\quad \widetilde{L}_kV_k^{(-)}=V_{k+1}^{(-)}L_k
\]
wlth the matrix}
\begin{equation}\label{dAL Vk-}
V_k^{(-)}=V_k^{(-)}(\wbq,\br,h)=\left(
\begin{array}{cc}1 & -h\lambda^{-1}\wiq_{k-1}\\ \\
-h\lambda^{-1} r_k & 1-h\lambda^{-2}+h\wiq_{k-1}r_k\end{array}\right).
\end{equation}

{\bf Proof} of this Proposition is completely parallel to that of the 
previous one and is therefore omitted.

{\bf Proposition 3.} {\it The Ablowitz--Ladik scheme {\rm (\ref{dAL})} with 
the parameters
\[
\alpha_0=\delta_0=\alpha_+=\delta_+=\delta_-=0, \quad \alpha_-=1
\]
is equivalent to the following map:
\begin{equation}\label{dAL flow- inv}
{\cal T}_-^{-1}(-h):\left\{\begin{array}{l}
(\wiq_k-q_k)/h=q_{k-1}(1-q_k\wir_k),\\ \\
(\wir_k-r_k)/h=-\wir_{k+1}(1-q_k\wir_k)\end{array}\right.
\end{equation}
approximating the flow ${\cal F}_-$. This map has the commutation 
representation
\[
{\cal T}_-^{-1}(-h):\quad \widetilde{L}_kW_k^{(-)}=W_{k+1}^{(-)}L_k
\]
with the matrix}
\begin{equation}\label{dAL Wk-}
W_k^{(-)}=W_k^{(-)}(\bq,\wbr,h)=\frac{1}{1-hq_{k-1}\wir_k}\left(
\begin{array}{cc}1+h\lambda^{-2}-h\wiq_{k-1}r_k & -h\lambda^{-1}q_{k-1}\\ \\
-h\lambda^{-1} \wir_k & 1\end{array}\right).
\end{equation}

{\bf Proof.} According to the Proposition 0, the matrix $W_k^{(-)}$ for the 
difference scheme ${\cal T}_{\rm AL}(h;0,0,1;0,0,0)$ has the form
\[
W_k^{(-)}=\left(
\begin{array}{cc}1+h\lambda^{-2}\Lambda_k & -h\lambda^{-1}q_{k-1}\Lambda_k\\ \\
-h\lambda^{-1}\wir_k\Lambda_k & 1+hq_{k-1}\wir_k\Lambda_k\end{array}\right).
\]
The equations of motion for this scheme may be presented as
\begin{equation}\label{proof3 aux1}
(\wiq_k-q_k)/h=q_{k-1}(1-\wiq_k\wir_k)\Lambda_k,\quad
(\wir_k-r_k)/h=-\wir_{k+1}(1-q_kr_k)\Lambda_{k+1}.
\end{equation}
Here $\Lambda_k$ is the solution of the difference equation
\begin{equation}\label{proof3 aux2}
\Lambda_{k+1}(1-q_kr_k)=\Lambda_k(1-\wiq_k\wir_k)
\end{equation}
tending to 1 by $k\to\pm\infty$ in the case of the rapidly decaying boundary
conditions. It is easy to check that the proposition will be proved if
we demonstrate the following formula for $\Lambda_k$:
\begin{equation}\label{proof3 aux3}
\Lambda_k=\frac{1}{1-hq_{k-1}\wir_k}
\end{equation}
which replaces in this case the general non-local expression (\ref{dAL Lamk}).
To do this, we first re-write (\ref{proof3 aux1}) as
\[
\wiq_k(1+hq_{k-1}\wir_k\Lambda_k)=q_k+hq_{k-1}\Lambda_k,\quad
r_k(1+hq_k\wir_{k+1}\Lambda_{k+1})=\wir_k+h\wir_{k+1}\Lambda_{k+1}.
\]
Now multiply the first of these equations by $\wir_k$, the second one by
$q_k$ and subtract the two resulting equations:
\[
(1-\wiq_k\wir_k)(1+hq_{k-1}\wir_k\Lambda_k)
=(1-q_kr_k)(1+hq_k\wir_{k+1}\Lambda_{k+1}).
\]
According to (\ref{proof3 aux2}), this is equivalent to
\[
\frac{\Lambda_{k+1}}{\Lambda_k}
=\frac{1+hq_k\wir_{k+1}\Lambda_{k+1}}{1+hq_{k-1}\wir_k\Lambda_k},
\]
or
\[
\frac{1}{\Lambda_k}+hq_{k-1}\wir_k=\frac{1}{\Lambda_{k+1}}+hq_k\wir_{k+1}
={\rm const.}
\]
Taking the $k\to\pm\infty$ limit, we see that this constant has to be equal
to 1, which finishes the proof.

{\bf Proposition 4.} {\it The Ablowitz--Ladik scheme {\rm (\ref{dAL})} with 
the parameters
\[
\alpha_0=\delta_0=\alpha_+=\delta_+=\alpha_-=0, \quad \delta_-=1
\]
is equivalent to the following map:
\begin{equation}\label{dAL flow+ inv}
{\cal T}_+^{-1}(-h):\left\{\begin{array}{l}
(\wiq_k-q_k)/h=\wiq_{k+1}(1-\wiq_kr_k),\\ \\
(\wir_k-r_k)/h=-r_{k-1}(1-\wiq_kr_k)\end{array}\right.
\end{equation}
approximating the flow ${\cal F}_+$. This map has the commutation 
representation
\[
{\cal T}_+^{-1}(-h):\quad \widetilde{L}_kW_k^{(+)}=W_{k+1}^{(+)}L_k
\]
with the matrix}
\begin{equation}\label{dAL Wk+}
W_k^{(+)}=W_k^{(+)}(\wbq,\br,h)=\frac{1}{1+h\wiq_kr_{k-1}}\left(
\begin{array}{cc}1 & h\lambda \wiq_k\\ \\
h\lambda r_{k-1} & 1-h\lambda^2+h\wiq_kr_{k-1}\end{array}\right).
\end{equation}

{\bf Proof} of this Proposition is omitted, because it is completely
analogous to that of the previous one.

{\bf Remark.}  It is important to notice the following relations:
\[
W_k^{(+)}(\wbq,\br,h)=(1-h\lambda^2)\Big(V_k^{(+)}(\wbq,\br,-h)\Big)^{-1},
\]
\[
W_k^{(-)}(\bq,\wbr,h)=(1+h\lambda^{-2})\Big(V_k^{(-)}(\bq,\wbr,-h)\Big)^{-1}.
\] 
These relations are very difficult to guess and to prove, if one remains by the original formulation
of the Proposition 0.

We have demonstrated so far that the maps ${\cal T}_+(h)$, ${\cal T}_-(h)$
have commutation representations with the same matrix $L_k$ as their
continuous time counterparts, and hence they share all the integrals of
motion. To claim that the maps belong to the Ablowitz--Ladik hierarchy,
we still need to show the Poisson property.

{\bf Proposition 5.} {\it The both maps ${\cal T}_+(h)$, ${\cal T}_-(h)$
are Poisson with respect to the Poisson bracket} (\ref{AL PB}).

{\bf Proof.} The Poisson property of a map $(\bq,\br)\mapsto(\wbq,\wbr)$
with respect to the bracket (\ref{AL PB}) is equivalent to the preservation 
of the corresponding 2-form:
\begin{equation}\label{proof5 aux1}
\sum_{k=1}^N\frac{d\wiq_k\wedge d\wir_k}{1-\wiq_k\wir_k}=
\sum_{k=1}^N\frac{dq_k\wedge dr_k}{1-q_kr_k}
\end{equation}
(remaining for simplicity by the finite dimensional case with the periodic
boundary conditions). We shall  prove this identity only for the map 
${\cal T}_+(h)$, since for ${\cal T}_-(h)$ everything is completely analogous. 
Differentiating the equations of motion (\ref{dAL flow+}), we obtain the 
following expressions:
\begin{equation}\label{proof5 aux2}
d\wiq_k=(1-hq_{k+1}\wir_k)\,dq_k+h(1-q_k\wir_k)\,dq_{k+1}-hq_{k+1}q_k\,d\wir_k,
\end{equation}
\begin{equation}\label{proof5 aux3}
(1-hq_k\wir_{k-1})\,d\wir_k=dr_k-h(1-q_k\wir_k)\,d\wir_{k-1}+h\wir_{k-1}\wir_k\,
dq_k.
\end{equation}
Using succesively these two formulas, we obtain:
\begin{eqnarray}
& &d\wiq_k\wedge d\wir_k= 
(1-hq_{k+1}\wir_k)\,dq_k\wedge d\wir_k+h(1-q_k\wir_k)\,dq_{k+1}\wedge d\wir_k
\nonumber\\
 &  & =\frac{1-hq_{k+1}\wir_k}{1-hq_k\wir_{k-1}}\,dq_k\wedge\Big(dr_k-
h(1-q_k\wir_k)\,d\wir_{k-1}\Big)+h(1-q_k\wir_k)\,dq_{k+1}\wedge d\wir_k.
\nonumber\\
\label{proof5 aux4}
\end{eqnarray}
To make the last step, we observe that the equations of motion (\ref{dAL flow+})
may be equivalently re-written as
\begin{equation}\label{proof10 aux2}
1-\wiq_k\wir_k=(1-q_k\wir_k)(1-hq_{k+1}\wir_k),
\end{equation}
\begin{equation}\label{proof10 aux3}
1-q_kr_k=(1-q_k\wir_k)(1-hq_k\wir_{k-1}).
\end{equation}
Upon use of (\ref{proof10 aux2}), (\ref{proof10 aux3}) the identity 
(\ref{proof5 aux4}) may be presented as
\[
\frac{d\wiq_k\wedge d\wir_k}{1-\wiq_k\wir_k}=
\frac{dq_k\wedge dr_k}{1-q_kr_k}
+\frac{h\,dq_{k+1}\wedge d\wir_k}{1-hq_{k+1}\wir_k}
-\frac{h\,dq_k\wedge d\wir_{k-1}}{1-hq_k\wir_{k-1}}.
\]
Summation over $k$ results in (\ref{proof5 aux1}). The proof is complete.

It should be mentioned that each of the pairs 
$({\cal T}_+(h),{\cal T}_-^{-1}(-h))$ and $({\cal T}_+^{-1}(-h),{\cal T}_-(h))$
may be seen as generated by one of the two simplest partitioned Runge--Kutta
methods \cite{HNW} when applied to the pair of differential systems 
$({\cal F}_+, {\cal F}_-)$. Recall that the both of these partitioned
Runge--Kutta methods are symplectic when applied to canonical Hamiltonian
systems \cite{SSS}. The Proposition 5 shows that this is still true for 
some systems which are Hamiltonian with respect to nonlinear Poisson brackets.
However, unlike the canonical case, this statement {\it cannot} be extended
to arbitrary systems Hamiltonian with respect to the bracket (\ref{AL PB}):
the concrete form of the right--hand side is essential for the validity of the
Proposition 5.

We finish this Section by constructing a discretization for the linear flow
${\cal F}_0$. This is a much more simple task.
Among many reasonable discretizations of the flow ${\cal F}_0$ we choose
(for the reasons which will become clear in the next Section)
\begin{equation}\label{dAL flow0}
{\cal T}_0(h):\quad
\wiq_k=\frac{1-h}{1+h}\,q_k,\quad \wir_k=\frac{1+h}{1-h}\,r_k.
\end{equation}

(Note that ${\cal T}_0^{-1}(-h)={\cal T}_0(h)$).

{\bf Proposition 6.} {\it The linear map ${\cal T}_0$ is Poisson with respect
to the bracket {\rm (\ref{AL PB})} and has the commutation 
representation  
\[
\widetilde{L}_kV^{(0)}=V^{(0)}L_k
\]
with the constant matrix}
\begin{equation}
V^{(0)}=V^{(0)}(h)=\left(
\begin{array}{cc}1-h & 0\\ \\0 & 1+h\end{array}\right).
\end{equation}

\setcounter{equation}{0}
\section{Local discretizations for ${\cal F}_0\circ{\cal F}_-\circ{\cal F}_+$}

Recall that, when considering the system (\ref{AL}) as a space 
discretization of the nonlinear Schr\"odinger equation (\ref{cNLS}), the
following reduction is of the primary interest:
\begin{equation}\label{AL red}
\br=\pm\bq^*,
\end{equation}
(it is admissible in the case of pure imaginary values of time $t$, that is, 
after the change of the independent variable $t\mapsto it$, $i=\sqrt{-1}$).

The flows ${\cal F}_+$, ${\cal F}_-$ alone do not allow this reduction any more, 
as well as their time discretizations ${\cal T}_+(h)$, ${\cal T}_-(h)$. 
Nevertheless, we shall demonstrate now that the composition of these maps
does again have this attractive property.

{\bf Proposition 7.} {\it The Ablowitz--Ladik scheme {\rm (\ref{dAL})} with 
the parameters
\[
\alpha_0=\delta_0=\alpha_-=\delta_-=0,\quad \alpha_+=\delta_+=1
\quad 
\]
may be presented as the composition}
\[
{\cal T}_-(h)\circ{\cal T}_+(h).
\]

{\bf Proposition 8.} {\it The Ablowitz--Ladik scheme {\rm (\ref{dAL})} with 
the parameters
\[
\alpha_0=\delta_0=\alpha_+=\delta_+=0,\quad \alpha_-=\delta_-=1
\quad 
\]
may be presented as the composition}
\[
{\cal T}_-(-h)^{-1}\circ{\cal T}_+(-h)^{-1}.
\]

{\bf Proof of the Proposition 7.} Let
\[
{\cal T}_+(h):\quad (\bq,\br)\mapsto (\widehat{\bq},\widehat{\br}),
\qquad {\cal T}_-(h):\quad (\widehat{\bq},\widehat{\br})\mapsto 
(\widetilde{\bq},\widetilde{\br}),
\]
so that, according to (\ref{dAL flow+}), (\ref{dAL flow-}),
\begin{equation}\label{T+ aux}
(\whq_k-q_k)/h=q_{k+1}(1-q_k\whr_k),\quad
(\whr_k-r_k)/h=-\whr_{k-1}(1-q_k\whr_k),
\end{equation}
\begin{equation}\label{T- aux}
(\wiq_k-\whq_k)/h=\wiq_{k-1}(1-\wiq_k\whr_k),\quad
(\wir_k-\whr_k)/h=-\whr_{k+1}(1-\wiq_k\whr_k).
\end{equation}
It follows from the Propositions 1,2 that the composition 
${\cal T}_-(h)\circ{\cal T}_+(h)$ allows the commutation representation
\begin{equation}
\widetilde{L}_kV_k=V_{k+1}L_k
\end{equation}
with the matrix 
\begin{equation}\label{Vk aux}
V_k=V_k^{(-)}(\widetilde{\bq},\widehat{\br})\,V_k^{(+)}(\bq,\widehat{\br}).
\end{equation}
We calculate now the entries of the matrix $V_k$ (denoting them according
to (\ref{dAL Vk})) in order to show that they have the form 
(\ref{dAL cAk})--(\ref{dAL cCk}). From (\ref{Vk aux}), (\ref{dAL Vk+}),
(\ref{dAL Vk-}) we obtain: 
\begin{eqnarray*}
{\cal B}_k & = & h\lambda q_k-h\lambda^{-1}\wiq_{k-1},\\
{\cal C}_k & = &-h\lambda^{-1}\whr_k(1+h\lambda^2-hq_k\whr_{k-1})+
h\lambda\whr_k(1-h\lambda^{-2}+h\wiq_{k-1}\whr_k)\\
 & = & h\lambda\Big(\whr_{k-1}-h\whr_k(1-\wiq_{k-1}\whr_{k-1})\Big)
-h\lambda^{-1}\Big(\whr_k+h\whr_{k-1}(1-q_k\whr_k)\Big)\\
 & = & h\lambda\wir_{k-1}-h\lambda^{-1}r_k
\end{eqnarray*}
(the last equality follows from the equations of motion (\ref{T+ aux}), 
(\ref{T- aux})),
\begin{eqnarray*}
{\cal A}_k & = & 1+h\lambda^2-hA_k \\
{\cal D}_k & = & 1-h\lambda^{-2}+hD_k,
\end{eqnarray*}
where
\begin{equation}\label{Ak Dk aux}
A_k=(q_k+h\wiq_{k-1})\whr_{k-1}, \qquad D_k=(\wiq_{k-1}-hq_k)\whr_k.
\end{equation}
So, the entries of the matrix $V_k$ has exactly the form 
(\ref{dAL cAk})--(\ref{dAL cCk}) with the parameters $\alpha_0=\delta_0=
\alpha_-=\delta_-=0$, $\alpha_+=\delta_+=1$. We may conclude
that the quantities $A_k$, $D_k$ satisfy the difference
relations (\ref{dif rel Ak}), (\ref{dif rel Dk}). (One could as well 
derive these difference relations (\ref{dif rel Ak}), (\ref{dif rel Dk}) 
directly from the definitions (\ref{Ak Dk aux}) and the equations of motion 
(\ref{T+ aux}), (\ref{T- aux}).) In the case of the rapidly decaying
boundary conditions we have obviously $A_k, D_k\to 0$ as $k\to\pm\infty$,
so that the quantities $A_k$, $D_k$ may be alternatively represented as in
(\ref{dAL Ak}), (\ref{dAL Dk}). This finishes the proof.

{\bf Proof of the Proposition 8.} This time let
\[
{\cal T}_+^{-1}(-h):\quad (\bq,\br)\mapsto (\widehat{\bq},\widehat{\br}),
\qquad {\cal T}_-^{-1}(-h):\quad (\widehat{\bq},\widehat{\br})\mapsto 
(\widetilde{\bq},\widetilde{\br}),
\]
so that, according to (\ref{dAL flow+ inv}), (\ref{dAL flow- inv}),
\begin{equation}\label{T+ inv aux}
(\whq_k-q_k)/h=\whq_{k+1}(1-\whq_kr_k),\quad
(\whr_k-r_k)/h=-r_{k-1}(1-\whq_kr_k),
\end{equation}
\begin{equation}\label{T- inv aux}
(\wiq_k-\whq_k)/h=\whq_{k-1}(1-\whq_k\wir_k),\quad
(\wir_k-\whr_k)/h=-\wir_{k+1}(1-\whq_k\wir_k).
\end{equation}
It follows from the Propositions 3,4 that the composition 
${\cal T}_-^{-1}(-h)\circ{\cal T}_+^{-1}(-h)$ allows the commutation 
representation
\begin{equation}
\widetilde{L}_kW_k=W_{k+1}L_k
\end{equation}
with the matrix 
\begin{equation}\label{Wk aux}
W_k=W_k^{(-)}(\widehat{\bq},\widetilde{\br})\,W_k^{(+)}(\widehat{\bq},\br).
\end{equation}
We calculate now the entries of the matrix $W_k$ (denoting them again according
to (\ref{dAL Vk})) in order to show that they, as in the previous case, have the 
form (\ref{dAL cAk})--(\ref{dAL cCk}). Denoting
\begin{equation}\label{Lamk aux}
\Lambda_k=\frac{1}{(1-h\whq_{k-1}\wir_k)(1+h\whq_kr_{k-1})},
\end{equation}
we obtain from (\ref{Wk aux}), (\ref{dAL Wk+}), (\ref{dAL Wk-}): 
\begin{eqnarray*}
{\cal C}_k\Lambda_k^{-1} & = & h\lambda r_{k-1}-h\lambda^{-1} \wir_k,\\
{\cal B}_k\Lambda_k^{-1} & = & h\lambda\whq_k(1+h\lambda^{-2}-h\whq_{k-1}\wir_k)
-h\lambda^{-1}\whq_{k-1}(1-h\lambda^2+h\whq_kr_{k-1})\\
 & = & h\lambda\Big(\whq_k+h\whq_{k-1}(1-\whq_k\wir_k)\Big)
-h\lambda^{-1}\Big(\whq_{k-1}-h\whq_k(1-\whq_{k-1}r_{k-1})\Big)\\
 & = & h\lambda\wiq_k-h\lambda^{-1}q_{k-1}
\end{eqnarray*}
(the last equality being based on the equations of motion (\ref{T+ inv aux}), 
(\ref{T- inv aux})),
\begin{eqnarray*}
{\cal A}_k\Lambda_k^{-1} & = & 
1-h\whq_{k-1}\wir_k-h^2\whq_{k-1}r_{k-1}+h\lambda^{-2}\\
& = & \Lambda_k^{-1}-
hr_{k-1}\Big(\whq_k+h\whq_{k-1}(1-\whq_k\wir_k)\Big)+h\lambda^{-2}\\
& = & \Lambda_k^{-1}-h\wiq_kr_{k-1}+h\lambda^{-2},\\
{\cal D}_k\Lambda_k^{-1} & = &
1+h\whq_kr_{k-1}-h^2\whq_k\wir_k-h\lambda^2\\
 & = & \Lambda_k^{-1}+
h\wir_k\Big(\whq_{k-1}-h\whq_k(1-\whq_{k-1}r_{k-1})\Big)-h\lambda^2\\
 & = & \Lambda_k^{-1}+hq_{k-1}\wir_k-h\lambda^2
\end{eqnarray*}
(using again repeatedly the equations of motion (\ref{T+ inv aux}), 
(\ref{T- inv aux})) and the definition (\ref{Lamk aux})). 

We see that this time the entries of the matrix $W_k$ have exactly the form 
(\ref{dAL cAk})--(\ref{dAL cCk}) with the parameters $\alpha_0=\delta_0=
\alpha_+=\delta_+=0$, $\alpha_-=\delta_-=1$. It follows that the 
quantity $\Lambda_k$ satisfies the difference relation (\ref{dif rel Lamk}) 
(which could be as well derived from the definition (\ref{Lamk aux}) and the 
equations of motion (\ref{T+ inv aux}), (\ref{T- inv aux}).) In the case of the 
rapidly decaying boundary conditions we have obviously $\Lambda_k\to 1$ as 
$k\to\pm\infty$, so that  $\Lambda_k$ may be alternatively represented as in
(\ref{dAL Lamk}). The proof is finished.

The maps constructed in these two Propositions have the desired property
(\ref{del = al conj}), assuring that the reduction (\ref{AL red}) is admissible
for them. However, they still do not approximate the Ablowitz--Ladik system
(\ref{AL}). In order to achieve this, we have to commute them with 
${\cal T}_0(h)$. The following lemma, following directly
from the formulas (\ref{dAL}), allows to control the parameters $\alpha_0$,
$\delta_0$ of the Ablowitz--Ladik discretizations.

{\bf Lemma.} {\it Let ${\cal T}'(h)={\cal T}_{\rm AL}(h;0,\alpha_+,\alpha_-;
0,\delta_+,\delta_-)$, and let ${\cal T}''(h)$ be the linear map
\[
(\bq,\br)\mapsto \left(\frac{1-h\alpha_0}{1+h\delta_0}\,\bq,
 \frac{1+h\delta_0}{1-h\alpha_0}\,\br\right).
\]
Then}
\begin{eqnarray*}
&&{\cal T}'(h)\circ{\cal T}''(h)={\cal T}''(h)\circ{\cal T}'(h)=\\
&&={\cal T}_{\rm AL}(h;\alpha_0,\alpha_+(1-h\alpha_0),\alpha_-(1-h\alpha_0);
\delta_0,\delta_+(1+h\delta_0),\delta_-(1+h\delta_0)).
\end{eqnarray*} 

According to this lemma, we derive from the Propositions 7,8 the following
fundamental statement.

{\bf Proposition 9.} {\it The Ablowitz--Ladik scheme {\rm (\ref{dAL})} with 
the parameters
\[
\alpha_0=\delta_0=1,\quad \alpha_+=1-h, \quad \delta_+=1+h,
\quad \alpha_-=\delta_-=0
\]
may be presented as the composition
\[
{\cal T}_0(h)\circ {\cal T}_-(h)\circ{\cal T}_+(h).
\]
The Ablowitz--Ladik scheme {\rm (\ref{dAL})} with the parameters
\[
\alpha_0=\delta_0=1,\quad \alpha_+=\delta_+=0, \quad \alpha_-=1-h, \quad
\delta_-=1+h
\]
may be presented as the composition}
\[
{\cal T}_0(h)\circ {\cal T}_-^{-1}(-h)\circ{\cal T}_+^{-1}(-h).
\]

These  two maps, approximating the system (\ref{dAL}), have the property
(\ref{del = al conj}) under the condition $h^*=-h$, i.e. $h$ pure 
imaginary. Hence they both may serve as honest time discretizations 
of the reduced version (\ref{AL real}) of the Ablowitz--Ladik system (\ref{AL}).
Note that all maps in each of these compositions commute.

\setcounter{equation}{0}
\section{Connection with relativistic Toda}

As noted in \cite{KMZ}, the Ablowitz--Ladik hierarchy is in principle
nothing but the relativistic Toda hierarchy and vice versa. We first 
give a Hamiltonian interpretation of this statement, and then use the 
results on the discrete time relativistic Toda lattice \cite{S2} to clarify
the place of the maps ${\cal T}_+$, ${\cal T}_-$ in the Ablowitz--Ladik
hierarchy.

Define new variables $c_k$, $d_k$ on the phase space of the Ablowitz--Ladik 
hierarchy:
\begin{equation}\label{qr to cd}
d_k=\frac{q_{k-1}}{q_k},\qquad c_k=\frac{q_{k-1}}{q_k}(q_kr_k-1).
\end{equation}
A direct computation shows that the only nonvanishing Poisson brackets 
between these functions are:
\begin{equation}\label{RTL PB}
\{c_k,d_{k+1}\}=c_kd_{k+1},\quad \{c_k,d_k\}=-c_kd_k,\quad 
\{c_k,c_{k+1}\}=c_kc_{k+1}.
\end{equation}
One immediately recognizes in these relations the quadratic Poisson brackets 
underlying the relativistic Toda hierarchy. Moreover, one sees immediately
that the simplest Hamiltonians of the Ablowitz--Ladik hierarchy $H_{\pm}$, $H_0$
may be expressed in the variables $c_k$, $d_k$ as
\begin{equation}\label{AL H+ to RTL}
H_+=\sum_{k=1}^N q_{k+1}r_k=\sum_{k=1}^N \frac{c_k+d_k}{d_kd_{k+1}},
\end{equation}
\begin{equation}\label{AL H- to RTL}
H_-=\sum_{k=1}^N q_kr_{k+1}=\sum_{k=1}^N (c_k+d_k),
\end{equation}
\begin{equation}\label{AL H0 to RTL}
H_0=\sum_{k=1}^N \log(1-q_kr_k)=\sum_{k=1}^N \log(c_k/d_k),
\end{equation}
and in $H_{\pm}$ we recognize the two basic Hamiltonians of the relativistic 
Toda hierarchy. We demonstrate now that our maps ${\cal T}_+$, ${\cal T}_-$,
being expressed in the variables $c_k$, $d_k$, also coincide with the
discrete time flows of the relativistic Toda lattice introduced in \cite{S2}.

{\bf Proposition 10.} {\it Consider the map ${\cal T}_+(h)$. Let the variables 
$c_k$, $d_k$ be defined by} (\ref{qr to cd}), {\it and define the auxiliary
function ${\goth d}_k$ by
\begin{equation}\label{T+ del}
{\goth d}_k=q_k\wir_k-1.
\end{equation}
Then the following relations hold:}
\begin{equation}\label{T+ recur}
\frac{c_k}{{\goth d}_k}=d_k-h-h{\goth d}_{k-1},
\end{equation}
\begin{equation}\label{T+ in cd}
\widetilde{d}_k=d_{k+1}\frac{d_k-h{\goth d}_{k-1}}{d_{k+1}-h{\goth d}_k},\quad
\widetilde{c}_k=c_{k+1}\frac{c_k+h{\goth d}_k}{c_{k+1}+h{\goth d}_{k+1}}.
\end{equation}

{\bf Proposition 11.} {\it Consider the map ${\cal T}_-(h)$. Let the variables 
$c_k$, $d_k$ be defined by} (\ref{qr to cd}), {\it and define the auxiliary
function ${\goth a}_k$ by
\begin{equation}\label{T- al}
{\goth a}_k=\frac{q_{k-1}}{\wiq_{k-1}}+\frac{hq_{k-1}}{q_k}.
\end{equation}
Then the following relations hold:}
\begin{equation}\label{T- recur}
{\goth a}_k=1+hd_k+\frac{hc_{k-1}}{{\goth a}_{k-1}},
\end{equation}
\begin{equation}\label{T- in cd}
\widetilde{d}_k=d_k\frac{{\goth a}_{k+1}-hd_{k+1}}{{\goth a}_k-hd_k},\quad
\widetilde{c}_k=c_k\frac{{\goth a}_{k+1}+hc_{k+1}}{{\goth a}_k+hc_k}.
\end{equation}

{\bf Proof of the Proposition 10}. We shall use in the proof the equations
of motion in the form (\ref{proof10 aux2}), (\ref{proof10 aux3}) and also
two additional auxiliary identities.
The first equation of motion in (\ref{dAL flow+}), re-written
with the help of definitions for $d_{k+1}$ and ${\goth d}_k$, reads:
\begin{equation}\label{proof10 aux1}
\frac{\wiq_k}{q_{k+1}}=d_{k+1}-h{\goth d}_k,
\end{equation}
Further, from (\ref{proof10 aux2}), (\ref{proof10 aux3}) and the definition 
(\ref{T+ del}) it follows:
\begin{equation}\label{proof10 aux4}
\frac{1-\wiq_k\wir_k}{1-q_{k+1}r_{k+1}}=\frac{{\goth d}_k}{{\goth d}_{k+1}}.
\end{equation}

Using the definitions (\ref{qr to cd}) for $c_k$ and (\ref{T+ del}) for 
${\goth d}_k$, then the identity (\ref{proof10 aux3}), and again the definitions 
(\ref{qr to cd}) for $d_k$ and (\ref{T+ del}) for ${\goth d}_{k-1}$, we get:
\[
\frac{c_k}{{\goth d}_k}=\frac{q_{k-1}}{q_k}\,\frac{1-q_kr_k}{1-q_k\wir_k}=
\frac{q_{k-1}}{q_k}\,(1-hq_k\wir_{k-1})=d_k-h-h{\goth d}_{k-1},
\]
which is the recurrent relation (\ref{T+ recur}).

To prove the first equality in (\ref{T+ in cd}), we use the definition of $d_k$
and the identity (\ref{proof10 aux1}): 
\[
\frac{\widetilde{d}_k}{d_{k+1}}=\frac{\wiq_{k-1}}{q_k}\,\frac{q_{k+1}}{\wiq_k}=
\frac{d_k-h{\goth d}_{k-1}}{d_{k+1}-h{\goth d}_k}.
\] 

Finally, to prove the second equality in (\ref{T+ in cd}), we use the
definition of $c_k$, the identities (\ref{proof10 aux1}), 
(\ref{proof10 aux4}), and the recurrent relation (\ref{T+ recur}):
\[
\frac{\widetilde{c}_k}{c_{k+1}}=\frac{\wiq_{k-1}}{q_k}\,\frac{q_{k+1}}{\wiq_k}\,
\frac{1-\wiq_k\wir_k}{1-q_{k+1}r_{k+1}}=
\frac{d_k-h{\goth d}_{k-1}}{d_{k+1}-h{\goth d}_k}\,
\frac{{\goth d}_k}{{\goth d}_{k+1}}=
\frac{c_k+h{\goth d}_k}{c_{k+1}+h{\goth d}_{k+1}}.
\]
The Proposition 10 is proved. 

{\bf Proof of the Proposition 11.} We start again with re-writing the equations
of motion (\ref{dAL flow-}) in the equivalent form:
\begin{equation}\label{proof11 aux3}
1-q_kr_k=(1-\wiq_kr_k)(1+h\wiq_{k-1}r_k),
\end{equation}
\begin{equation}\label{proof11 aux4}
1-\wiq_k\wir_k=(1-\wiq_kr_k)(1+h\wiq_kr_{k+1}).
\end{equation}
Note that the first equation of motion in (\ref{dAL flow-}) may be represented
also in another equivalent form:
\begin{equation}\label{proof11 aux5}
\wiq_k(1+h\wiq_{k-1}r_k)=q_k+h\wiq_{k-1}.
\end{equation}
We shall need also two additional auxiliary identities. The definitions 
(\ref{T- al}), (\ref{qr to cd}) immediately imply:
\begin{equation}\label{proof11 aux1}
{\goth a}_k-hd_k=\frac{q_{k-1}}{\wiq_{k-1}},
\end{equation}
\begin{equation}\label{proof11 aux2}
{\goth a}_k+hc_k=\frac{q_{k-1}}{\wiq_{k-1}}(1+h\wiq_{k-1}r_k).
\end{equation}

Now to prove the recurrent relation (\ref{T- recur}) we use (\ref{proof11 aux2})
in conjunction with (\ref{T- al}), and then (\ref{proof11 aux5}) and
(\ref{proof11 aux1}):
\[
1+\frac{hc_k}{{\goth a}_k}=\frac{q_k(1+h\wiq_{k-1}r_k)}{q_k+h\wiq_{k-1}}=
\frac{q_k}{\wiq_k}={\goth a}_{k+1}-hd_{k+1}.
\]
The first equation of motion in (\ref{T- in cd}) follows from the definition 
of $d_k$ and the formula (\ref{proof11 aux1}):
\[
\frac{\widetilde{d}_k}{d_k}=\frac{\wiq_{k-1}}{q_{k-1}}\,\frac{q_k}{\wiq_k}=
\frac{{\goth a}_{k+1}-hd_{k+1}}{{\goth a}_k-hd_k}.
\]
Finally, to prove the second equation of motion in (\ref{T- in cd}), we use
the definition of $c_k$, the formulas (\ref{proof11 aux3}), (\ref{proof11 aux4}),
and then (\ref{proof11 aux5}):
\[
\frac{\widetilde{c}_k}{c_k}=\frac{\wiq_{k-1}}{q_{k-1}}\,\frac{q_k}{\wiq_k}\,
\frac{1-\wiq_k\wir_k}{1-q_kr_k}=
\frac{\wiq_{k-1}}{q_{k-1}}\,\frac{q_k}{\wiq_k}\,
\frac{1+h\wiq_kr_{k+1}}{1+h\wiq_{k-1}r_k}=
\frac{{\goth a}_{k+1}+hc_{k+1}}{{\goth a}_k+hc_k}.
\]
The Proposition 11 is proved.

Now one immediately recognizes in (\ref{T+ recur}), (\ref{T+ in cd}) and
in (\ref{T- recur}), (\ref{T- in cd}) two discrete time flows of the 
relativistic Toda hierarchy introduced and studied in \cite{S2}. Applying 
the results in \cite{S2}, we get the following statement (formulated, as 
before, for the periodic case for the sake of notational simplicity).

{\bf Proposition 12.} {\it The maps ${\cal T}_{\pm}$ have Lax representations
with either of the $N\times N$ Lax matrices
\[
T_+(\bq,\br,\lambda)=L(\bq,\br,\lambda)U^{-1}(\bq,\br,\lambda)\quad {\rm or}
\quad T_-(\bq,\br,\lambda)=U^{-1}(\bq,\br,\lambda)L(\bq,\br,\lambda),
\]
where
\[
L(\bq,\br,\lambda)=\sum_{k=1}^N \frac{q_{k-1}}{q_k}E_{kk}+\lambda\sum_{k=1}^N
E_{k+1,k},
\]
\[
U(\bq,\br,\lambda)=\sum_{k=1}^N E_{kk}+\lambda^{-1}\sum_{k=1}^N
\frac{q_{k-1}}{q_k}(1-q_kr_k)E_{k,k+1}.
\]
These maps are interpolated by the flows with the Hamiltonian functions
\[
-{\rm tr}_0(\Phi(-T_{\pm}^{-1}))=H_++O(h) \qquad {\rm and}\qquad 
{\rm tr}_0(\Phi(T_{\pm}))=H_-+O(h),
\]
respectively, where
\[
\Phi(\xi)=h^{-1}\int_0^{\xi}\log(1+h\eta)\frac{d\eta}{\eta}=\xi+O(h).
\]
The initial value problems for the maps ${\cal T}_{\pm}$ may be solved in
terms of the matrix factorization problem for the matrices
\[
\Big(I-hT^{-1}_{\pm}(t=0)\Big)^n\qquad {\rm and} \qquad 
\Big(I+hT_{\pm}(t=0)\Big)^n,
\]
respectively.}

(In the formulation above ${\rm tr}_0$ stands for the free term of the Laurent
expansion for the trace; the detailed definition of the matrix factorization 
problem in a loop group mentioned in this Proposition, may be found in 
\cite{S2}).

\section{Conclusion}
Re-considering the Ablowitz--Ladik discretizations from the modern point
of view, undertaken in this paper, turned out to be rather fruitful. 
We factorized a highly non-local scheme into the product of very
simple (local) ones, each of them approximating a more simple and
fundamental flow of the Ablowitz--Ladik hierarchy. These local schemes
may be stuided exhaustively. In particular, we found in this paper the
interpolating Hamiltonian flows for them, as well as the solution in terms
of factorization problem in a loop group. We guess that also in the
practical computations our variant of the difference scheme will exceed
considerably the old one. It would be interesting and important to carry 
out the corresponding numerical experiments.

It seems also promising to re-consider from this point of view other
non-local integrable discretizations derived and tested in \cite{AT3}.

We note also that our maps are ideal building blocks for applying
the Ruth--Yoshida--Suzuki techniques \cite{RYS}, which will result in higher 
order integrable discretizations for the Ablowitz--Ladik system. This point 
will be reported in detail elsewhere.

The research of the author is financially supported by the DFG (Deutsche
Forschungsgemeinshaft).

\end{document}